\documentstyle[prl,aps]{revtex}

\begin{document}

\title{Two-phonon scattering of magnetorotons in fractional 
 quantum Hall liquids}

\author{V.M. Apalkov and M.E. Portnoi\cite{byline}}
\address{School of Physics, University of Exeter, Stocker Road, 
 Exeter EX4 4QL, United Kingdom}
%\date{\today}
\maketitle
%\voffset=2.0true cm

\begin{abstract}
We study the phonon-assisted process of dissociation 
of a magnetoroton, in a fractional quantum Hall liquid, 
into an unbound pair of quasiparticles. Whilst the 
dissociation is forbidden to first order in the electron-phonon 
interaction, it can occur as a two-phonon process. 
Depending on the value of final separation between 
the quasiparticles, the dissociation is either 
a single event involving absorption of one phonon and 
emission of another phonon of similar energy,
or a two-phonon diffusion of a quasiexciton in momentum space.
The dependence of the magnetoroton dissociation time on the 
filling factor of the incompressible liquid is found.
\end{abstract}

 One of the successful methods used to study the properties 
 of two-dimensional (2D) electron systems is phonon absorption 
 spectroscopy\cite{phonon0,phonon1,phonon2}.
 In this  method the non-equilibrium acoustic phonons injected into the system 
 are absorbed by the 2D electron gas. Such a technique has recently been 
 used to acquire the information about the magnetically quantized 
 2D electron gas at fractional filling factors\cite{phonon2}. 
 The absorption of the non-equilibrium phonons by a 2D incompressible 
 liquid, underlying the fractional quantum Hall effect, results in 
 the increase of an effective electron temperature, which 
 can be detected by measuring the dissipative conductivity of the system.

 The phonon absorption process can be schematically 
 divided into two stages. During the first stage,  
 the phonons are absorbed by the incompressible liquid, 
 with creation of the non-equilibrium neutral excitations
 (magnetorotons) with a wave vector of the order of the inverse 
 magnetic length\cite{benedict}.
 During the second stage, the non-equilibrium electron 
 system is thermalized, with the creation of charged  quasiparticles 
 which are free to participate in the dissipative conductivity. 
 The process of thermalization 
 can be understood as the dissociation of magnetorotons into 
 well-separated quasielectrons and quasiholes. 
 Initially, it was proposed that such a process could be intensified 
 by absorption of non-equilibrium phonons 
 from the pulse\cite{benedict1}. It was later shown\cite{benedict2}
 that due to the almost flat dispersion of the neutral 
 excitations of an incompressible liquid at large wave vectors, 
 the absorption of a phonon by a magnetoroton is allowed only 
 for a very small momentum transfer, which is not enough to dissociate 
 the magnetoroton. In the present paper we show that the phonon-pulse 
 induced dissociation of the magnetoroton occurs as a second-order 
 process in the electron-phonon interaction. 
 It is a common feature for all systems containing dispersionless particles, 
 that the two-phonon processes provide the main contribution 
 to the transport properties of the system\cite{bychkov,falko}.

 The neutral low-energy excitations of the incompressible electron 
 liquid have a non-zero gap $\Delta (q)$ for all wave vectors $q$,  
 with a minimum value $\Delta ^{*}$ at $q=q^*\sim 1/l_0$, 
 where $l_0$ is the magnetic length.
 The excitations close to $q^*$ are well-described by the single mode 
 approximation\cite{GMP}, and are called magnetorotons in analogy to 
 roton excitations of superfluid helium. At large wave vectors the 
 neutral excitation is a quasiexciton, which consists of the fractionally 
 charged quasielectron and quasihole\cite{GMP,QHE0,QHE1,QHE2}. In what 
 follows, we consider quantum Hall liquids with a filling factor 
 $\nu = p/m$, where $p$ is an integer and $m$ is an odd number. 
 For this filling factor the charge of a quasielectron 
 (quasihole) is $1/m$ $(-1/m)$ of the electron charge. 
 The effective magnetic length for these quasiparticles 
 is $~l_0^{*}=\sqrt{m}l_0$.
 The dispersion of neutral excitations of the incompressible 
 liquid at large values of momentum can then be written in the form: 
 $\Delta (q) =  \Delta _{\infty } - \varepsilon _0/m^3 q$, 
 where $\varepsilon _0 = e^2/\kappa l_0$ is the Coulomb energy. 
 We use the magnetic length $l_0$ and the Coulomb 
 energy $\varepsilon _0$ as the units of length and energy, respectively. 
 To study the phonon-induced dissociation of a magnetoroton we 
 should consider the phonon-assisted transitions of the magnetoroton 
 with momentum close to $q^*$ into the quasiexciton with momentum 
 larger than a critical momentum value $q_{cr}$. This critical momentum 
 is defined by the condition:
\begin{equation}
\Delta_{\infty} - \Delta(q_{cr}) = T_{cr}  \mbox{\hspace{3mm},}
\label{eq1}
\end{equation}
 where $T_{cr}$ is a characteristic temperature (in units of 
 $\varepsilon_0/k_B$). 

 To find the rate of two-phonon scattering of neutral 
 excitations of an incompressible liquid,  we assume that 
 within the whole range of our interest ($q>q^{*}$), a magnetoroton
 can be described as a quasiexciton, i.e., a bound state of 
 a quasielectron and a quasihole.  
 Although this is not a very good  approximation at 
 $q\sim q^{*}$, we use this assumption to illustrate the importance  
 of two-phonon processes for the thermal relaxation of 
 incompressible liquids. 

 For convenience we label the three-dimensional (3D) vectors with
 capital letters, e.g., $\vec{Q}$, and their  projections  
 with the corresponding lower-case letters: $\vec{Q}=(\vec{q},q_z)$.
 We neglect the effects related to phonon polarization and 
 crystal anisotropy, and write down the  quasielectron-phonon 
 and quasihole-phonon interaction Hamiltonian in the form:
\begin{equation}
H_{int}  =  \sum _{\vec{Q}} \frac{1}{m}\frac{V_{e}(Q)}{\sqrt{\Omega }} Z(q_z) 
         \hat{\rho} _{e} (\vec{q}) \left[ 
           \hat{b}^{+} (-\vec{Q}) + 
           \hat{b} (\vec{Q}) \right]   
    - \sum _{\vec{Q}} \frac{1}{m} \frac{V_{e}(Q)}{\sqrt{\Omega }} Z(q_z) 
           \hat{\rho }_{h}(\vec{q}) \left[ 
          \hat{b}^{+} (-\vec{Q}) + 
           \hat{b} (\vec{Q}) \right]~,  
\label{eq2}
\end{equation}
 where $\hat{b}^{+}(\vec{Q}),~\hat{b}(\vec{Q})$ are creation and 
 annihilation operators of a phonon with momentum $\vec{Q}$, 
 $\Omega $ is a normalization volume, and $\hat{\rho }_{e}$ and 
 $\hat{\rho } _{h}$ are the quasielectron and quasihole 2D density 
 operators, respectively. Because the quasiparticles are in the 
 lowest subband of size quantization, the $z$-part of the density
 operator is accumulated in the form-factor $Z(q_z)$:
\begin{equation}
Z(q_z) = \frac{1}{(1+iq_z/\beta )^3} \mbox{\hspace{3mm},}
\end{equation}
 where the Fang-Howard approximation\cite{ando} for the lowest 
 subband wave function is used, and $\beta $ is the 
 parameter of the Fang-Howard wave function. 

 The response of a quasiparticle to the external perturbation
 is $m$ times weaker than the corresponding response of an
 electron. For the quasihole this response has the opposite sign. 
 Therefore, in Eq.~(\ref{eq2})  the quasiparticle-phonon 
 interaction is characterized by the matrix elements 
 $\pm \frac{1}{m} V_{e}(Q)$, where
 the upper sign is for quasielectrons and the lower sign is
 for quasiholes. Here, $V_{e}(Q)$ is the matrix element of the 
 electron-phonon interaction, which is given by 
 the expression\cite{levinson}:
\begin{equation}
 V_{e}(Q)  = \sqrt{\frac{B_{p}}{Q}} - i \sqrt{B_{d} Q}
             \mbox{\hspace{3mm},}
\label{eq3}
\end{equation}
 where $B_p$, $B_q$ are the constants of piezoelectric and deformation 
 potential couplings. We use material parameters of GaAs in our 
 numerical estimates.  

 Using Eq.~(\ref{eq3}) one can rewrite the Hamiltonian 
 (\ref{eq1}) in the form:
\begin{equation}
H_{int}  =  \sum _{\vec{Q}} \frac{1}{m} \frac{ Z(q_z)}{\sqrt{\Omega }} 
          \left( \sqrt{\frac{B_{p}}{Q}}   
      - i \sqrt{B_{d} Q} \right)  \hat{c}(\vec{q}) 
         \left[  \hat{b}^{+} (-\vec{Q}) +  \hat{b} (\vec{Q})     \right]
                    \mbox{\hspace{3mm},}
\label{eq4}
\end{equation}
where 
$
\hat{c}(\vec{q}) =  \hat{\rho} _{e} (\vec{q}) - 
         \hat{\rho} _{h} (\vec{q})
$ is the quasiparticle charge density operator.

 The matrix elements of the operator $\hat{c}(\vec{q})$ 
 between the quasiexciton states, which are characterized by the 2D 
 momenta $\vec{k}$ and $\vec{k}_1$, are given by the expression:
\begin{equation}
 \left< \vec{k}_1 \right| \hat{c}(\vec{q}) \left| \vec{k} \right>
  = -2i e^{-mq^2/4 } \sin\left(m\frac{q_x k_y - q_y k_x}{2 } \right) 
   \delta(\vec{k}_1 -\vec{k} -\vec{q})  
  \label{eq5}  
\end{equation}
where the momentum is in units of $1/l_0$. 

 The rate of two-phonon scattering of the quasiexciton 
 to states with momentum greater than $q_{cr}$ is:
\begin{eqnarray}
W(q_{cr}) & = & \frac{2 \pi}{\hbar \varepsilon _0 ^3 l_0^6} \frac{1}{m^4}
  \int \frac{d\varphi _{q^*}}{2\pi }  \int _{q_f>q_{cr}} d^2 \vec{q}_f 
   \int \int \frac{d^3 \vec{Q}}{(2\pi )^3} 
             \frac{d^3 \vec{Q^{\prime }}}{(2\pi )^3} 
  |Z(q_z)|^2 |Z(q^{\prime }_z)|^2  |M(\vec{q}^*,\vec{q}_f,\vec{Q},
\vec{Q^{\prime }})|^2                         \times  \nonumber \\
    &  &  \times  N(Q)(N(Q^{\prime })+1) \:
\delta \! \left( \Delta(q^{*})-\Delta(q_f) - s(Q^{\prime }-Q) \right)  \:
\delta \! \left( \vec{q}^{*}+\vec{q}-\vec{q}_f-\vec{q}^{\prime} \right)
                        \mbox{\hspace{3mm},}
\label{eq7}
\end{eqnarray}
where 
\begin{eqnarray}
M(\vec{q}^*,\vec{q}_f,\vec{Q},\vec{Q^{\prime }}) & = & 
\frac{ <\vec{q}^*| \hat{H}(\vec{Q}) |\vec{q}^* + \vec{q}>
       <\vec{q}^* + \vec{q}|\hat{H}(\vec{Q^{\prime }}) | \vec{q}_f >}
     { \Delta_a -s Q^{\prime }} 
                                  \nonumber  \\
 & & + \frac{ <\vec{q}^*| \hat{H}(\vec{Q^{\prime }}) 
                         |\vec{q}^* + \vec{q}^{\prime }>
       <\vec{q}^* + \vec{q}^{\prime }|\hat{H}(\vec{Q}) | \vec{q}_f >}
     { \Delta_b +s Q^{\prime }} \mbox{\hspace{3mm},} 
\label{eq8}
\end{eqnarray}
$\Delta_a = \Delta(\vec{q}^* +\vec{q}) - \Delta(\vec{q}_f) $ , 
$\Delta_b = \Delta(\vec{q}^* +\vec{q}^{\prime }) - \Delta(\vec{q}^*)~$
 and
\begin{equation}
\hat{H}(\vec{Q}) = \left[ \sqrt{\frac{B_{p}l_0}{Q}}   
      - i \sqrt{\frac{B_{d}Q}{l_0}} \right] 
                        \hat{c}(\vec{q})   \mbox{\hspace{3mm},}
\label{eq9}
\end{equation}
 where $s$ is the average speed of sound, which is $s\approx 0.03$ 
 in units of $\varepsilon _0 l_0/\hbar$. In Eq.~(\ref{eq7}) the 
 first integral is the average of the scattering rate over all 
 possible directions of the initial momentum $\vec{q}^{*}$. $N(Q)$ 
 is the non-equilibrium phonon distribution function, created 
 by the external phonon source. The two delta-functions represent 
 the conservation of energy and momentum. 

 Due to the exponential dependence (Eq.~(\ref{eq5})) of the matrix 
 elements of the charge density operator $\hat{c}(\vec{q})$ 
 on the phonon momentum $q$, the main contribution 
 to the integrals in Eq.~(\ref{eq7}) comes from 
 the region where the difference between $q$ and $ q^{\prime }$ 
 is small, and the vectors $\vec{q}^*$ and $\vec{q}_f$ 
 are parallel. In this case, substituting Eqs.~(\ref{eq5}) 
 and (\ref{eq9}) into Eq.~(\ref{eq8}), we obtain the expression:
%\begin{eqnarray}
%|M(\vec{q}^*,\vec{q}_f,\vec{Q},\vec{Q^{\prime }})|^2  &  \approx &
%16 \left[ \frac{B_p l_0}{Q}+ 
%          \frac{B_d Q}{l_0} \right]
%   \left[  \frac{B_p l_0}{Q^{\prime }} + 
%           \frac{B_d Q^{\prime }}{l_0} \right]
%\frac{1}{(\Delta _1 (q_f) +s Q^{\prime })^2}  \nonumber \\
% & \times & \left\{
% \left[ \frac{\delta \Delta }{\Delta _1 (q_f) +s Q^{\prime }} \right]^2
% \left[ \sin \frac{mqq^* \sin \phi}{2 } \right]^4
% \left[ \cos \frac{mqq^{\prime } \sin 2\phi}{2 } \right]^2 \right. 
%                       \nonumber \\ 
% & & + \left. \left[ \sin \frac{mqq^{\prime }\sin 2\phi }{2 }  
%        \sin \frac{mqq_f \sin 2\psi }{2 }  \right]^2 \right\}     
%      \mbox{\hspace{3mm},}
%\label{eq10}
%\end{eqnarray}
\begin{eqnarray}
|M(\vec{q}^*,\vec{q}_f,\vec{Q},\vec{Q^{\prime }})|^2 & \approx &
16 \frac{  ( B_p l_0^2 + B_d Q^2)( B_p l_0^2 + B_d Q^{\prime 2})}
        {QQ^{\prime} l_0^2 (\Delta _1 (q_f) +s Q^{\prime })^2}
       \left\{
 \left[ \frac{\delta \Delta }{\Delta _1 (q_f) +s Q^{\prime }} \right]^2
 \left[ \sin \frac{mqq^* \sin \phi}{2 } \right]^4
        \right. 
\nonumber\\
& & \times \left.
 \left[ \cos \frac{mqq^{\prime } \sin 2\phi}{2 } \right]^2  
      + \left[ \sin \frac{mqq^{\prime }\sin 2\phi }{2}  
        \sin \frac{mqq_f \sin 2\psi }{2 }  \right]^2 \right\} ~,     
\label{eq10}
\end{eqnarray}
where $~\Delta _1(q_{cr}) \approx \Delta_a \approx -\Delta_b \approx 
 \Delta((q_{cr}+q^*)/2)-\Delta(q^{*})$,
  $~\delta \Delta  = \Delta _a +\Delta _b \approx 2 
 \Delta(\vec{q}^* +\vec{q}) -
 \Delta(\vec{q}_f) - \Delta(\vec{q}^*)$, $\phi $ is the angle between
 vectors $\vec{q}$ and $\vec{q}^{*}$, and $\psi $ is the angle between
 vectors $\vec{q}^*$ and $\vec{q}_f$.  From Eq.(\ref{eq10}) one can see
 that the amplitude of two-phonon scattering of the quasiexciton  
 vanishes when the wave vectors of the phonons, 
 $\vec{q}$ and $\vec{q}^{\prime}$,
 have the same direction. We assume that the angles $\phi $ and
 $\psi $ between the wave vectors are small. This approximation is
 good for large values of $q_f$. 
  The first term in the curly brackets in Eq.~(\ref{eq10}) 
 contains the factor $\delta \Delta $,
 which describes the nonlinearity of the excitation spectra and 
 depends strongly on the shape of the dispersion curve $\Delta (q)$. 
 There is no good approximation which describes the magnetoroton 
 spectrum in the intermediate region between the magnetoroton minimum 
 and the beginning of the quasiexciton-like dispersion.
 As the upper limit for $~\delta \Delta ~$ we take the value 
$\delta \Delta = 0.005$, which corresponds to a quarter of the 
 magnetoroton binding energy;
 the term containing $\delta \Delta $ 
 then gives only a small correction to the whole expression.  
 In what follows we, shall disregard this term and only consider
 the contribution from the second term in the curly brackets. 
 
 If the width of the wave function in $z$-direction is much larger than 
 $2/\delta q _{cr}$, i.e., the Fang-Howard parameter $\beta$ is 
 much smaller than $\delta q _{cr}/2$, then by 
 substituting Eq.~(\ref{eq10}) into Eq.~(\ref{eq7}) and performing the 
 integrations we obtain the expression for the rate of quasiexciton 
 dissociation:
\begin{eqnarray}
W(q_{cr}) & \approx &  W_0 \frac{1}{m^4}\frac{3}{2 \pi^2 }
    \frac{ 1}{s^3} \frac{q_{cr}\sqrt{q_{cr}}}{\sqrt{q^*}}
\left[ \frac{\beta^2 s}{\delta \! q_{cr} ~\Delta_0(q_{cr} )}
     \right]^{7/2} 
  N(\delta q_{cr}/2) (N(\delta q_{cr}/2)+1) \nonumber \\
  & & \times
 \left[  1+ \frac{B_d ~\delta \! q_{cr}^2}{4B_p l_0^2} \right]^2
 \frac{1}{\delta \! q_{cr}}\left(1+3 \frac{q_{cr}}{(q^*+q_{cr})^2}\right)
 \mbox{erfc} 
 \left(\frac{\sqrt{m}}{2}\delta \! q_{cr} \right)  \mbox{\hspace{3mm},}
\label{eq11}
\end{eqnarray}
 where $\delta \! q_{cr} = q_{cr} - q^*$, $\mbox{erfc}(x)= 
1-\mbox{erf}(x)=\frac{2}{\sqrt{\pi}} \int_{x}^{\infty } 
dy\exp(-y^2)$ is the complementary error function,  and the constant
$W_0$ is given by
$ W_0 = B_{p}^2 / (\hbar \varepsilon _0^3 l_0^4) $.

In the opposite case of a narrow quantum well, 
i.e., $\beta \gg \delta q_{cr}/2$, $~W(q_{cr})$ takes the form:
\begin{eqnarray}
W(q_{cr}) & \approx & W_0 \frac{1}{m^4} \frac{1 }{6 \pi^3 }
    \frac{ 1}{s^3}  \frac{q_{cr}\sqrt{q_{cr}}}{\sqrt{q^*}}
 N(\delta q_{cr}/2)( N(\delta q_{cr}/2)+1)  
                                                           \nonumber \\
 & \times & \frac{1}{\delta \! q_{cr}}
 \left(1+3 \frac{q_{cr}}{(q^*+q_{cr})^2}\right)
 \mbox{erfc} 
 \left(\frac{\sqrt{m}}{2}\delta \! q_{cr} \right)  \mbox{\hspace{3mm},}
\label{eq12}
\end{eqnarray}
where the main contribution comes from the piezoelectric interaction.

 From Eqs.~(\ref{eq11}) and (\ref{eq12}) one can see that the rate
 $W(q_{cr})$ is exponentially small for large  $q_{cr}$. 
The two-phonon dissociation
of the quasiexciton as a single event is allowed only for small 
values of critical momentum. For example,  for $q_{cr} \approx 3$,  
 which corresponds to a characteristic temperature larger than 
 $T_{cr}\sim 1/m^3 q_{cr}\sim 1.5$K (see 
 Eq.(\ref{eq1})), we find from Eq.~(\ref{eq12}) the rate of quasiexciton 
 dissociation 
$W(q_{cr}) \approx 3\times 10^8 ~$s$^{-1}$, or the characteristic
time of dissociation $\sim 3 \times 10^{-9}~$s. For this calculation
 the temperature of non-equilibrium phonons is taken as $T_{ph}=10$K.
For a greater value of the critical momentum $q_{cr}$, the rate of  
 direct two-phonon dissociation of quasiexciton becomes very small. For 
example, at $T_{cr}=0.5$K we obtain $q_{cr}\sim 1/m ^3 T_{cr} \sim 7$, 
which gives an exponentially small value of the erfc function: 
$\mbox{erfc} \left(\frac{\sqrt{3}}{2}(q_{cr}-q^*) \right)\sim 10^{-16}$.
 In this case one should consider the series of subsequent two-phonon 
 transitions, which represent the diffusion of the quasiexciton in 
 momentum space. 

Following the derivation of Eqs.~(\ref{eq11}) and (\ref{eq12}) we 
estimate the average change in the  momentum of the 
quasiexciton, $\delta q_0$, during a two-phonon 
scattering event: $\delta q_0 \approx 2/\sqrt{m}$, which is 
equal to 1.15 for $\nu =1/3$.

 The time of the two-phonon scattering event
  can be estimated from Eq.~(\ref{eq7}), where 
 the lower integration limit is taken as $q^*$ ($q_f>q^*$). 
 For a narrow quantum well this time is 
\begin{equation}
\tau _{0}  \approx   \frac{1}{W_0} 6\pi^3 m^4 
    \frac{s^3 }{N(\delta q_{0}/2)( N(\delta q_{0}/2)+1)}
  \frac{\sqrt{q^*}}{(q^*+\delta q_0)^{3/2}}~ \delta q_0 
 \left(1+3 \frac{q_{cr}}{(q^*+q_{cr})^2}\right)^{-1}
  \mbox{\hspace{3mm}.}
\label{eq14}
\end{equation}
The time of diffusion of the quasiexciton from the magnetoroton
 state with momentum $q^*$ to the quasiexciton state with  
critical momentum $q_{cr}$ can then be estimated from the equation:
\begin{equation}
\tau _{cr} \approx \tau _0 \left( \frac {q_{cr}}{\delta q_0} \right)^2 \approx 
 \tau_0 \frac{1}{m^6} \left( \frac{1}{T_{cr} \delta q_0} \right)^2
                     \mbox{\hspace{3mm},}
\label{eq15}
\end{equation}
where Eq.~(\ref{eq1}) is used. 

 For $\nu =1/3$, $s = 0.03$,  and for the non-equilibrium phonon temperature 
 $T_{ph}=10~$K our estimate gives 
 $\tau _0 \approx 0.4\times 10^{-10}~$s. In this
 case the time of the magnetoroton dissociation is about $\tau _{cr} 
 \approx 2\times   10^{-9}~$s for $T_{cr}=0.5$K. 

 We assume that the phonon distribution is characterized by some 
 equilibrium temperature $T_{ph}$ of the phonon pulse: 
 $N(q)=(e^{sq/T_{ph}}-1)^{-1}$. The typical phonon energy 
 in Eq.~(\ref{eq14}) is about $s\delta q_0/2=s/\sqrt{m}$, which corresponds
 to a temperature $T_{ph}^* \approx s/\sqrt{m}$. At $\nu =1/3$ we
 have $T_{ph}^*\approx 2$K. If the phonon non-equilibrium 
 temperature is much larger than $T_{ph}^*$, $T_{ph}\gg T_{ph}^*$, 
 then the dependence of $\tau _0$  on the denominator $m$ of 
 the filling factor has the power law $\tau _0 \propto m^{7/2}$.
 From Eq.~(\ref{eq15}), the time of magnetoroton dissociation 
 $\tau _{cr}$ is then proportional to $ m^{-3/2}$. 

 In conclusion, we have shown that a flux of non-equilibrium phonons 
 induces dissociation of magnetorotons of an incompressible 
 fractional quantum Hall liquid into well-separated pairs of 
 quasiparticles.  This process is only possible  
 to second order in the quasiparticle-phonon interaction. 
 Depending on the value of the critical quasiexciton momentum, the 
 magnetoroton decay can be considered as a direct two-phonon 
 dissociation or as the result of the two-phonon diffusion  
 of a quasiexciton in momentum space. 

 This work was supported by the UK EPSRC.

\end{document}